\documentclass[conference]{IEEEtran}
\ifCLASSINFOpdf
 \usepackage[pdftex]{graphicx}
\else
\fi
\usepackage{url}

\usepackage{listings}
\usepackage{color}
\definecolor{javared}{rgb}{0.6,0,0} 
\definecolor{javagreen}{rgb}{0.25,0.5,0.35} 
\definecolor{javapurple}{rgb}{0.5,0,0.35} 
\definecolor{javadocblue}{rgb}{0.25,0.35,0.75} 
 
\begin{document}
%
\title{Towards an Enterprise-Ready Implementation of Artificial Intelligence-Enabled, Blockchain-Based Smart Contracts}

\author{\IEEEauthorblockN{Philipp Brune\IEEEauthorrefmark{1}\IEEEauthorrefmark{2}}
\IEEEauthorblockA{\IEEEauthorrefmark{1}Department of Information Management\\
Neu-Ulm University of Applied Sciences\\
Wileystra{\ss}e 1, 89231 Neu-Ulm, Germany\\
Email:Philipp.Brune@hs-neu-ulm.de}
\IEEEauthorblockA{\IEEEauthorrefmark{2}QWICS Enterprise Systems\\
Taunustor 1, 60310 Frankfurt am Main, Germany\\
Email:Philipp.Brune@qwicschain.com}
}


%


\maketitle

\begin{abstract}
Blockchain technology and artificial intelligence (AI) are current hot topics in research and practice. However, the potentials of their combination have been studied just recently to a larger extend. While different use cases for combining AI and blockchain have been discussed, the idea of enabling blockchain-based smart contracts to perform ``smarter'' decisions by using AI or machine learning (ML) models has only been considered on the conceptual level so far. It remained open, how such AI-enabled smart contracts could be implemented in a robust way for real-world applications. Therefore, in this paper a new, enterprise-class implementation of AI-enabled smart contracts is presented and first insights regarding its feasibility are discussed.
\end{abstract}


%
\IEEEpeerreviewmaketitle

\section{Introduction}

The concepts of blockchain and smart contracts became extremely popular in recent years, emerging from Bitcoin \cite{Nakamoto2008} and the cryptocurrency hype it induced, even though many of its ideas existed in the literature before \cite{Lamport1982,Szabo1997}.

The combination of blockchain \--- or (better) Distributed Ledger Technology (DLT) \--- applications \cite{Wuest2018} with Artificial Intelligence (AI) is a rather new topic, which just started to receive more attention in recent years \cite{Corea2017,Telcoin2018,Lopes2018}. Possible applications for this combination have been grouped in the following categories \cite{Salah2019}: (1) Decentralized data storage and management with AI, (2) DLT as a decentralized infrastructure for AI, and (3) decentralized AI applications \cite{Salah2019}.

While for (1), AI or machine learning (ML) models are implemented off-chain and just access the data stored on the DLT, a scenario suggested e.g. for medical and healthcare applications \cite{Lopes2018,Salah2019}, (2) and (3) focus on actually executing at least parts of the AI/ML processes on-chain in a distributed manner \cite{Mylrea2018,Harris2019}. 

Some approaches for using DLT as a platform for distributed training or deployment of AI/ML models have also been proposed in practice, e.g. ChainIntel \cite{Chainintel2018} or SingularityNET\footnote{\url{https://singularitynet.io/}}. This is were smart contracts come into play, for executing either the training or inference phases of ML models in a distributed manner on the nodes of the blockchain network.

Smart contracts have been proposed already years ago \cite{Szabo1997} as computer protocols which represent the terms and conditions of a legal contract in the form of executable program code \cite{VonHaller2016}. In combination with Blockchain, they have gained a lot of attention in recent years as a tool for building so-called decentralized applications (DApps), e.g. on the Ethereum blockchain \cite{Wood2014}. While ``smarter'' smart contracts have been discussed previously with respect to increase their security \cite{Luu2016}, to make them really ``smart'', they should be enabled to use AI-based prediction models for their decisions. Potential use cases for such AI-enabled smart contracts could e.g. be to handle anti-money laundering or fraud management on-chain or to manage volatility in cryptocurrencies \cite{Telcoin2018}. Other authors suggest their use for the next-generation energy IoT infrastructure \cite{Mylrea2018}.

However, these previous publications discuss the integration of ML models into smart contracts either as a general research challenge, and thus on a rather high conceptual level, or for very specific use cases only. In particular, it has not been discussed so far, how such AI-enabled smart contracts could be implemented in a robust way for real-world enterprise applications. Therefore, in this paper the question is addressed, how an enterprise-class implementation of AI-enabled smart contracts could be achieved, which addresses also the needs of traditional IT operations, e.g. in the financial services industry.

The rest of this paper is organized as follows: In section \ref{arch} the overall software architecture of the approach is described. Section \ref{example} illustrates the actual embedding of ML models in the smart contracts by example. We conclude with a summary of our findings.

\section{Software Architecture}
\label{arch}

\begin{figure*}[!t]
\centering
\includegraphics[width=5.5in]{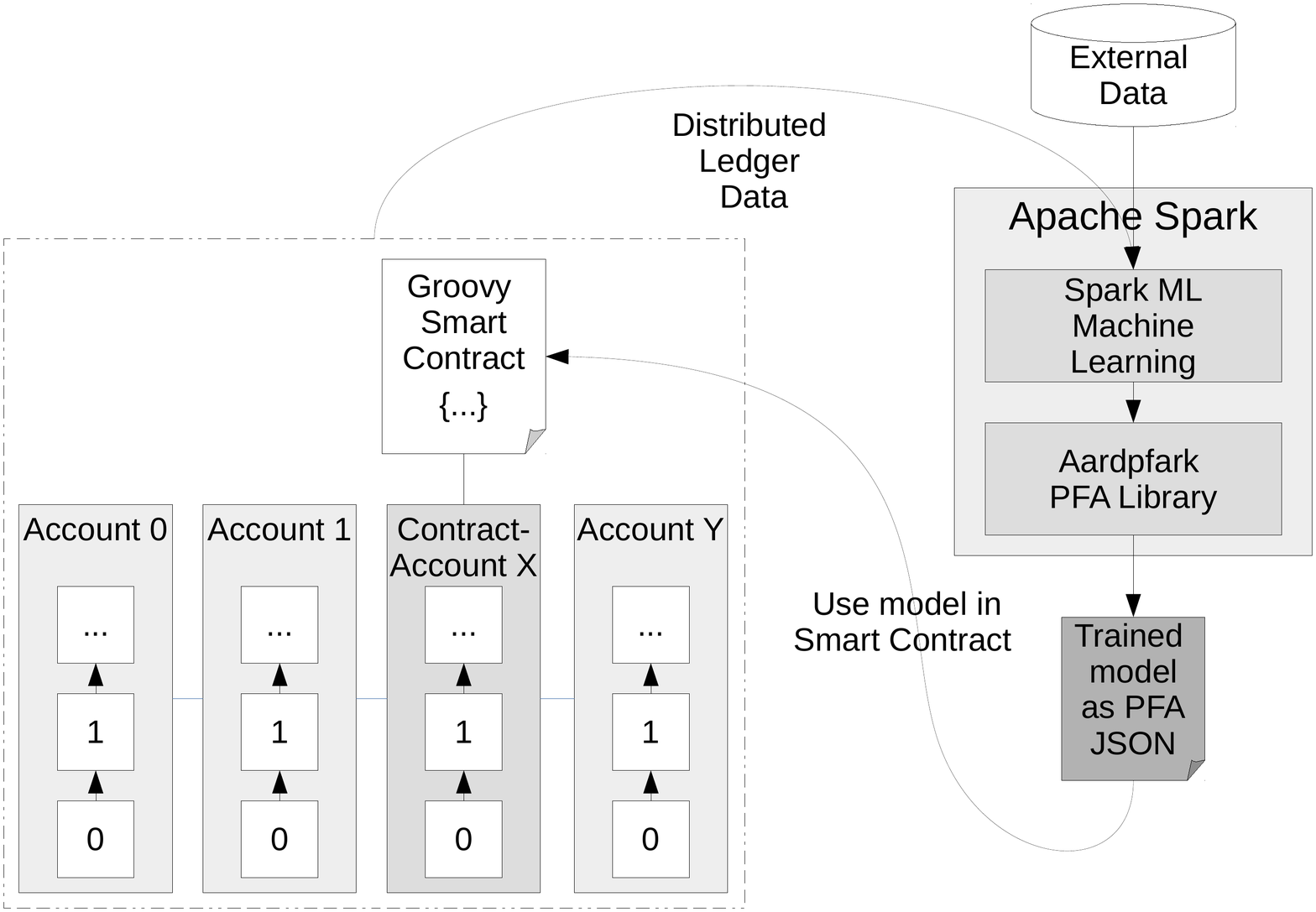}
\caption{Overall process of training and using ML models in combination with the QWICSchain DLT framework and its Groovy-based smart contract engine schematically shown left \cite{Brune2019}. Predictive models could be trained by external tools like e.g. Apache Spark based on data from the blockchain as well as from other exteranl sources. The Portable Format for Analytics (PFA) standard is used to distribute the trained models.}
\label{proc}
\end{figure*}

As the basis for the proposed proof-of-concept (PoC), the open-source QWICSchain blockchain framework\footnote{\url{https://github.com/pbrune1973/qwicschain}} was selected, as it already provides a more traditional, enterprise-class blockchain implementation built on Java Enterprise Edition(EE)\footnote{Recently, Java EE has been handed over to the Eclipse Foundation to manage its future development, and therefore re-labeled as Jakarta EE (\url{https://jakarta.ee/about/}). However, for the sake of simplicity, in this paper still only the term Java EE is used to denote both Java EE and Jakarta EE.} \cite{Brune2019}. 

This framework uses the Java Virtual Machine( JVM)-based Apache Groovy language\footnote{\url{http://groovy-lang.org/}} to implement smart contracts. 
In recent years, the Groovy scripting language became considerably popular among developrs, e.g. in the financial services industry, in particular due to its easy handling of big decimal numbers \cite{Taft2013}. 

The proposed solution addresses the use of ML models trained by various platforms and tools outside the blockchain (off-chain), potentially analyzing both the chain data itself as well as any kind of external data sources. These trained models are then used inside the smart contracts running on the blockchain (on-chain) to score new feature values (so-called inference). 

This process is illustrated in figure \ref{proc}. Here, as an example, the Apache Spark framework and its machine learning (ML) library\footnote{\url{https://spark.apache.org/}} are used to train the models. 

To represent the trained models, the approach uses the Portable Format for Analytics (PFA) standard \footnote{\url{http://dmg.org/pfa/}}, which enables data scientists to distribute of ML models as JavaScript Object Notation (JSON) documents. With Apache Spark, the open-source Aardpfark library\footnote{\url{https://github.com/CODAIT/aardpfark}} could be used to export the trained models as PFA documents. To execute these models inside Groovy on the JVM, the open-source Hadrian PFA implementation library\footnote{\url{https://github.com/modelop/hadrian}} was selected.

\begin{figure}[!t]
\centering
\includegraphics[width=3in]{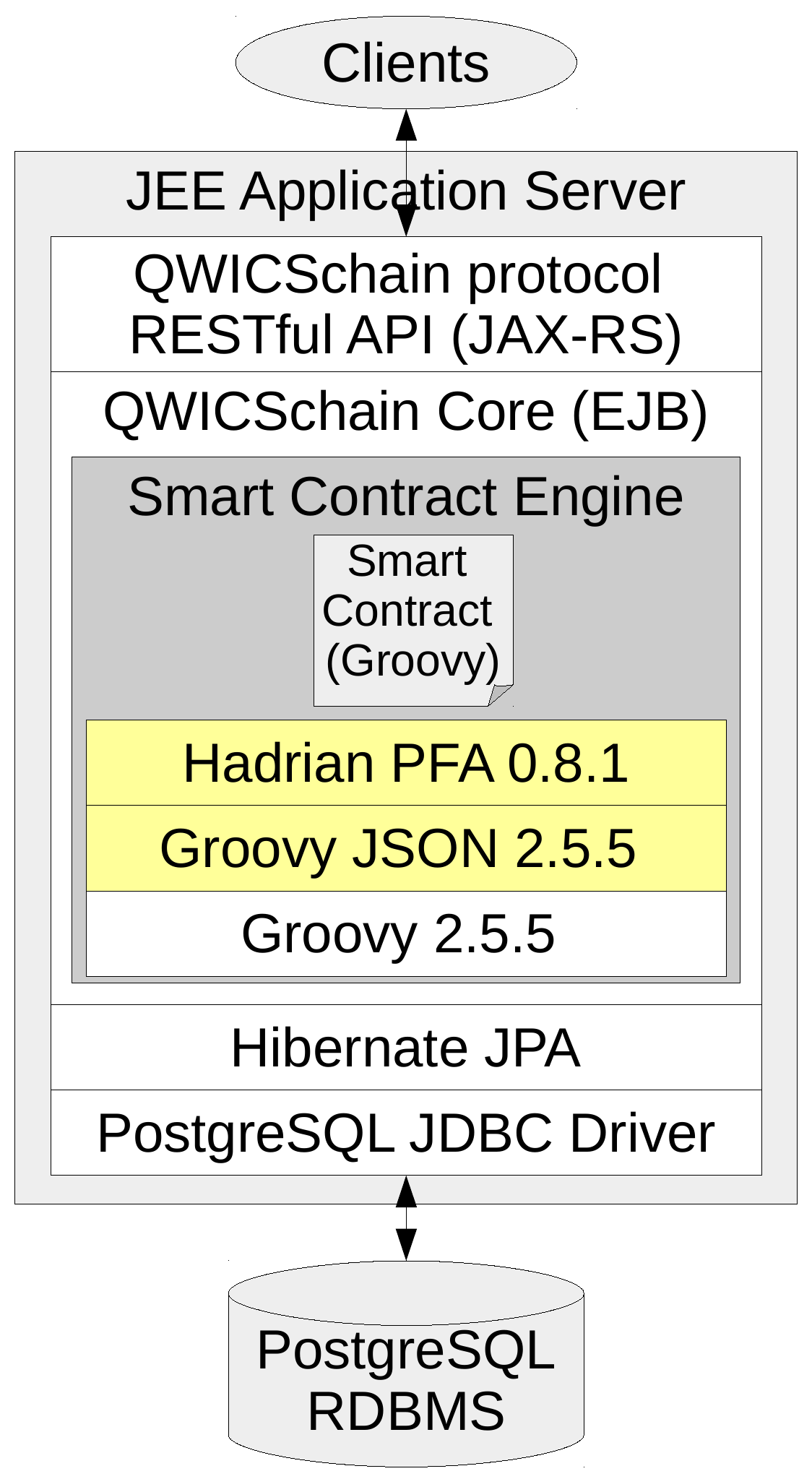}
\caption{Extensions (yellow boxes) of the original QWICSchain architecture(white and grey boxes) \cite{Brune2019} for the usage of trained ML models inside its Groovy-based smart contracts.}
\label{archi}
\end{figure}

Figure \ref{archi} shows the software architecture of proposed approach, extending the original QWICSchain framework \cite{Brune2019} by AI support. The existing Groovy-based smart contract implementation is now extended to incorporate the use of trained ML models as part of the smart contracts code. Therefore, the respective model definitions are deployed as part of the smart contract source code on the blockchain, to ensure full transparency and immutability of these models as well.

\section{Smart Contract Examples}
\label{example}

In QWICSchain \cite{Brune2019}, smart contracts are written in the Groovy language. Every smart contract resides on its dedicated contract account on the distributed ledger. This account may receive or send transactions and is fully controlled by its smart contract. This structure is illustrated schematically in figure \ref{proc} as well. 

Therefore, smart contracts act like automated account owners. This behaviour is partially similar to other blockchain frameworks, e.g. Ethereum \cite{Wood2014}. However, different from the latter, smart contracts are written in an established scripting language here, not a proprietary notation like Solidity. 

Smart contracts may change the state of the ledger only through sending transactions to other accounts. They are allowed to read the complete state of the ledger, but any change to it may performed only through transaction sending.

\begin{lstlisting}[caption=Basic structure of a smart contract written in Groovy for the QWICSchain framrework.,label=lst1]
class MyContract extends Contract {
	...
	void receive(Long senderId, 
		Integer action, BigDecimal coins, 
		Asset asset, byte[] data) {
		...
		// Send an outgoing transaction
		send(receiverId,...) 
		...
	}
}
\end{lstlisting}

In listing \ref{lst1}, the basic structure of such a smart contract is shown. It is represented by a class (here called {\it MyContract}) inheriting from the framework class {\it Contract}. At minimum, it has to implement the inherited method {\it receive(...)}, which is called by the framework everytime a valid transaction is received by and booked on the contract account.

As mentioned above, during execution of this method, the code may now change the state of the ledger by sending transactions to arbitrary receiving accounts via calling the inherited method {\it send(...)}.

Additionally, every smart contract has a dedicated, persistent storage for maintaining its state accross multiple transactions. It is implemented as a key-value store for objects of any kind, with the keys being string identifiers.

This basic structure has now been extended for the present work to incorporate the use of trained PFA models inside such smart contracts. Therefore, two new methods have been added to the framework's {\it Contract} class (written in Java) as outlined in listing \ref{lst2}.

Here, the first method creates a model instance from a textual description. The argument {\it lang} denotes the formal language, in which the model is represented. Currently, the only allowed value is ``PFA'', but this parameter allows for future extensions to other model representations. The second parameter is a string value representing the model (e.g. in JSON notation).

\begin{lstlisting}[caption=Structure of the extension of the existing framework's Java class {\it Contract} by two methods {\it createModel(...)} and {\it score()...)} to support the use PFA-based ML models inside smart contracts,label=lst2]
public class Contract {
	...	
	protected final Object createModel(
		String lang, String def) {
		...
	}	

	protected final Object score(
		Object model, Object data) {
		...
	}
}
\end{lstlisting}

Listing \ref{lst3} illustrates, how these functions could used inside an example smart contract class called {\it MyContract}. The model definition in PFA format is stored as a string value in the attribute variable {\it modelJson}.

\begin{lstlisting}[caption=Groovy code fragment of a smart contract using a trained ML model. The PFA JSON representation of the model is stored in the attribute variable {\it modelJson}.,label=lst3]
class MyContract extends Contract {
	def modelJson = '''{
  		"name":"mlpc_204d01de546b",
 		"version":1,
 		"doc":"Auto-generated by ...",
 		 ...
 	}'''	 

	void receive(Long senderId, 
		Integer action, BigDecimal coins, 
		Asset asset, byte[] data) {
		...
		// Example of using a PFA model
		def input = new FeatureData(
			[-0.166667,0.416667,
			-0.0169491,-0.0833333])
		def model = createModel("PFA",
			modelJson)
		log(input.toString())
		def sc = score(model,input)
		log(""+sc.prediction)		
		...
	}
}
\end{lstlisting}

Inside the overridden {\it receive(...)} method, first an examplified feature data set object is created, which should be scored. The framework provides the wrapper class {\it FeatureData} for this purpose, of which the constructor takes an array object as a parameter, representing the data vector to be scored. Subsequently, the model is created using the inherited {\it createModel(...)} method and stored in the local variable {\it model}.

Then the  {\it score(...)} method is called, passing it the  {\it model} and  {\it input} objects to perform the model inference and calculate the prediction score value {\it sc} for the input data.

The inherited method {\it log(...)} is used for development purposes only, printing its parameter value as a string to the developer's console.

This approach allows to use a a trained prediction model inside a smart contract, and ensures the model and its parameters are stored in a readable way on the blockchain. This makes all decisions made by a contract using such kind of models transparent and auditable by all participants of the network.

So far, the solution has been implemented as illustrated in figures \ref{proc} and \ref{arch} and evaluated by using a simple test model. The on-chain functionality is now already included in the QWICSchain online demo environment available at \url{https://qwicschain.com} for further testing.

\section{Conclusion}

In conclusion, in this paper an approach has been presented, how an enterprise-class implementation of AI-enabled smart contracts could be realized, which uses well-established standards such as the Java EE platform, the Groovy scripting language for wiriting smart contracts, and the standardized PFA representation of ML models. 

The general handling and usage of trained ML models as part of the code of smart contracts has been demonstrated, which is intended to enable AI-based ``smart'' decisions on the blockchain. 

While the approach looks promising, further research is needed to evaluate it in lab and field tests and develop and roll out practically relevant use cases for such a combination of AI with blockchain technology.

\bibliographystyle{IEEEtran}
\bibliography{IEEEabrv,refs.bib}

\end{document}